\documentclass[12pt,onecolumn, journal, draftcls]{IEEEtran}

\usepackage{amsmath}
\usepackage{epsfig}
\usepackage{amssymb}
\usepackage{graphicx}
\usepackage{color}
\usepackage{cite}

\newtheorem{thry}{\textbf{Theorem}}
\begin{document}
\title{Power Allocation Strategies across $N$ Orthogonal Channels at Both Source and Relay} 

\author{
\IEEEauthorblockN{
Youngwook~Ko,~\IEEEmembership{Member,~IEEE,}
Masoud~Ardakani,~\IEEEmembership{Senior Member,~IEEE,}
and~Sergiy~A.~Vorobyov,~\IEEEmembership{Senior Member,~IEEE}
}
\thanks{The authors are with the Department
of Electrical and Computer Engineering, University of Alberta,
Edmonton, AB, T6G~2V4 Canada, e-mails:
(\{ko,~ardakani,~vorobyov\}@ece.ualberta.ca).}
\thanks{This work is supported in part by the Natural Science
and Engineering Research Council (NSERC) of Canada and the Alberta
Ingenuity Foundation, Alberta, Canada.}
}

\maketitle

\vspace{-1.5cm}

\baselineskip .78cm

\begin{abstract}
We consider a wireless relay network with one source, one relay and
one destination, where communications between nodes are preformed
via $N$ orthogonal channels. This, for example, is the case when
orthogonal frequency division multiplexing is employed for
data communications. Since the power available at the source and
relay is limited, we study optimal power allocation strategies at
the source and relay in order to maximize the overall
source-destination capacity under individual power
constraints at the source and/or the relay. Depending on the availability
of the channel state information at the source and rely, optimal
power allocation strategies are performed at both the source and
relay or only at the relay. Considering different setups for the
problem, various optimization problems are formulated and solved.
Some properties of the optimal solution are also proved.
\end{abstract}



\baselineskip .83cm

\section{Introduction}

Driven by the ever increasing demand for high--speed services within limited spectrum resources, cooperative relay transmission, as a promising
spectrally efficient technique, has received significant
interests in recent years \cite{sendonaris_tc03:aazhang, laneman_it04, karmar_ti05:gupta, peters_mcom09:heath}. Cooperative relaying promises numerous gains for wireless networks such as improved reliability \cite{laneman_it04} and increased network capacity \cite{karmar_ti05:gupta}. The benefits of relay transmission can also be reaped by orthogonal frequency
division multiple (OFDM) access systems to support broadband
services, e.g., the IEEE~802.16j
\cite{peters_mcom09:heath}.

Power limitation is a common problem in wireless networks. Therefore, allocating the
limited power resources to the network nodes (i.e., the source and relay nodes) is a design
consideration which has received much attention \cite{bletsas_wc07, tannious_jsac08, hwang_wc08, guo_globecom08:li, cui_it09:poor, madsen_ti06:zhang, hasna_tw04:alouini, zhao_tw07:lim, quek_icc07:win, pham_ts10:tuan, chen_tw08:serbetli, panah_tv09:heath, serbetli_tc08:yener, PLVT08conf, phan_tw09:vorobyov, phan_EURASIP09:vorobyov, zhou_icc09:lin}. For example, relay selection schemes, where only a subset of relay nodes is considered for cooperation, are studied in \cite{bletsas_wc07, tannious_jsac08, hwang_wc08, guo_globecom08:li, cui_it09:poor} as a simple power allocation technique.
It has been shown in \cite{madsen_ti06:zhang} that the optimal power allocation
between the source and relay nodes can improve the network performance.
Moreover, considering a multi-hop network in
\cite{hasna_tw04:alouini}, the optimal allocation of power among the
hops is shown to significantly improve the performance.
Also, in \cite{zhao_tw07:lim, quek_icc07:win, pham_ts10:tuan, chen_tw08:serbetli}, assuming that a
common message is sent from the source node to multiple amplify--and--forward (AF) or decode--and--forward (DF) relay nodes over
orthogonal channels, optimal power allocation strategies among
relay nodes are studied.


Recently, considering relay networks with multiple
source-destination pairs (multicast), relevant power allocation
strategies have been addressed \cite{panah_tv09:heath, PLVT08conf,
phan_tw09:vorobyov, phan_EURASIP09:vorobyov, serbetli_tc08:yener,
zhou_icc09:lin}. The set up of \cite{panah_tv09:heath} is that one
source sends independent messages to a number of destinations. A
single OFDM channel is used between the source and a single relay
node, where each private message is assigned to one frequency tone
and then amplified and forwarded on different frequency tones to
various users. Thus, power allocation at the relay is studied to
maximize the minimum SNR among all destinations. In
\cite{PLVT08conf, phan_tw09:vorobyov, phan_EURASIP09:vorobyov},
various power allocation schemes at both the source and relay
nodes are developed for the general multi-source, -relay, and
-destination set up. Transmissions are done over orthogonal
channels, and the corresponding power allocation schemes are based
on maximizing the minimum SNR among all users, minimizing the
maximum transmit power over all sources, as well as maximizing the
network throughput, the minimum rate among all users, and the
weighted-sum of rates. Power allocation at the relay nodes only in
the multi-relay multi-destination set up is considered also in
\cite{serbetli_tc08:yener}, and is based on maximizing the sum
capacity of the network. In \cite{zhou_icc09:lin}, power
allocation at both the source and relay nodes in a set up
similar to \cite{phan_tw09:vorobyov, phan_EURASIP09:vorobyov} is
studied for maximizing the sum network capacity under power
constraints for orthogonal subchannels.

In this paper, we consider a two--hop AF relay network, which
consists of a single source node, a single destination, and a single relay nodes. We assume that the
communications between nodes occur across $N$ orthogonal subchannels, where each subchannel is assigned independent
information. Thus, the network capacity is the sum of the individual
capacities of subchannels. This setup may correspond to the case, for example, when OFDM signalling is employed
between the nodes. Our motivation for considering this setup is the widespread integration of OFDM into
various wireless standards. Therefore, power allocation among the orthogonal channels (across frequency tones) is an important issue.

Taking into account individual power constraints at the source and
relay nodes, we study strategies for optimal allocation of the
limited power among the $N$ orthogonal subchannels. Depending on the
availability of channel state information (CSI) at the source and/or
the relay, the power allocation is done at both the source node and the relay node or only at the relay node. The goal is to maximize the overall data rate of the network.

We consider two cases for data forwarding at the relay. In one case, we assume the information received on the $i$-th source--relay subchannel is amplified and forwarded on the $i$-th relay-destination subchannel. In the other case, similar to \cite{hottinen_ciss06}, we allow the relay to switch the source message received on one source--relay subchannel to another relay--destination subchannel. With optimal power allocation, this strategy significantly improves the overall achievable rate.

Power allocation for a similar OFDM--based relay network has been studied in \cite{hammerstrom_tw07:wittneben}. The main differences of our work are as follows.
(i) With global CSI available at both the source and relay, we first prove that the optimal solution should satisfy a certain symmetry. This symmetry, in turn, allows us to directly find the optimal power allocation for both the source and relay. In contrast, in \cite{hammerstrom_tw07:wittneben}, an iterative optimization approach is studied. An iterative solution may not find the global optima in a reasonable number of iterations. Our approach, however, finds the optimal solution in one shot.
(ii) We also consider the situation when only the relay has global CSI. In other words, the source only knows the source-relay channels. This is a more practical setup as the source need not know the relay-destination channels. For this case, we propose a simple greedy algorithm to maximize the overall network achievable rate. It is also worth noting here that the optimization method of \cite{hammerstrom_tw07:wittneben} needs global CSI and cannot be used for the latter case.

Note that availability of global CSI (and
therefore optimal power allocation) only at the relay has been
considered in, for example, \cite{serbetli_tc08:yener}.
However, the existing optimal power allocation schemes at the relay mostly focus on maximizing the sum capacity and it is not clear how the source (who lacks global CSI) should allocate rates across subchannels.
Instead, our approach maximizes the achievable sum rate, while
considering the outage. For this, we study how the source chooses its actual data rates across subchannels to avoid outage.
Interestingly, at high SNR, the achievable
rate is very close to that obtained in the case when global CSI is available to both
source and relay.

The rest of this paper is organized as follows. In Section
\ref{sec:sys}, we outline the system model. Our optimization
problems for maximizing the achievable sum rate are also defined. In
Section~\ref{sec:opt}, we propose optimal power allocation
strategies at both the source and relay or only at the
relay for different cases with respect to the availability of
CSI. Some of the properties of the optimal solution are also studied.
Section \ref{sec:asym} provides asymptotic analysis for optimal power allocation at high SNR. In Section \ref{sec:sim}, numerical results
are provided to demonstrate the benefits of the optimal power
allocation. The paper is concluded in Section \ref{sec:con}.

\section{System model and problem definitions}
\label{sec:sys}

\subsection{System and channel models}

Consider a two--hop relay network which consists of one source node ($S$), one destination node ($D$), and one relay node ($R$) with no line of sight between $S$ and $D.$ Thus, $R$ is employed in order to assist the communications between $S$ and $D.$ Moreover, it is supposed that communication between nodes is done accross $N$ orthogonal links.

Two cases are considered for data forwarding at $R$. In the first case, the message received on the $i$-th $S$ to $R$ subchannel ($SR_i$) is simply amplified and forwarded by the relay on the $i$-th $R$ to $D$ subchannel ($RD_i$). This case is referred to as AF relaying. In the second case, as in \cite{hottinen_ciss06}, we assume that the relay sorts $S$ to $R$ subchannels  and $R$ to $D$ subchannels based on their quality and assigns data on the $i$-th best $S$ to $R$ subchannel to the $i$-th best $R$ to $D$ subchannel. This amplify--sort--and--forward setup is referred to hereafter as ASF. For both AF and ASF cases, complete transmissions from $S$ to $D$ are composed of two phases: (i) $S$ sends the messages while $R$ listens; (ii) $R$ forwards the message to $D$.

Considering that each node is equipped with a single antenna, the received signal at $R$ through $SR_i$ subchannel can be expressed as
\begin{equation}
\label{eq:io_phase1}
y_{i} = g_i \sqrt{P_{S_i}} m_i + n_i
\end{equation}
where $m_i$ is the source message with a unit energy transmitted by $S$ on the link $SR_i,$ $P_{S_i}$ is the transmit power on the link $SR_i,$ $g_i$ is the associated flat Rayleigh fading channel coefficient for this link, which is a complex Gaussian random variable with zero mean and variance $\sigma_{g_i}^2,$ i.e., $g_i \sim {\cal CN}(0,\sigma_{g_i}^2),$ and $n_i$ stands for a complex valued additive white Gaussian noise (AWGN) such that $n_i \sim {\cal CN}(0,1).$

For AF relaying, at the end of the second phase, the received signal at $D$ via subchannel $RD_i$ can be written as
\begin{equation}
\label{eq:io_phase2:af}
x_{i} = h_i \sqrt{P_{R_i}} \frac{y_i}{ \sqrt{ E\{ |y_i|^2 \} } } + n'_i
\end{equation}
where $E\{\cdot\}$ denotes the expected operation, $P_{R_i}$ is the transmit power
assigned to the link $RD_i,$ $h_i \sim {\cal
CN}(0,\sigma^2_{h_i})$ is the channel coefficient for this link, and $n'_i \sim {\cal CN}(0,1)$ is the AWGN.

Moreover, for the ASF relay network, the received signal at $D$ from $R$ via the $i$-th ordered link (denoted as $RD_{(i)}$) can be represented as
\begin{equation}
\label{eq:io_phase2:asf}
x_{(i)} = h_{(i)} \sqrt{P_{R_{(i)}}} \frac{y_{(i)}}{ \sqrt{ E\{ |y_{(i)}|^2 \} } } + n'_{(i)}
\end{equation}
where the subscript $(i)$ stands for the $i$-th ordered link when sorting is based on $|h_i|$ in the decreasing order such that $|h_{(1)}|\geq \cdots \geq |h_{(N)}|.$ Also, $P_{R_{(i)}}$ is the transmit power on the link $RD_{(i)},$ and $n'_{(i)} \sim {\cal CN}(0,1)$ is the AWGN. Note that $y_{(i)}$ in (\ref{eq:io_phase2:asf}) can be found from (\ref{eq:io_phase1}) when all $SR_i$ links are ordered to $SR_{(i)}$ and again sorted based on $|g_i|$ in the decreasing order such that $|g_{(1)}|\geq \cdots \geq |g_{(N)}|.$

This system can be viewed as $N$ subsystems operating on separate orthogonal suchchannels. For the $i$-th subsystem, we can define a source, a relay, and a destination denoted by $S_i,$ $R_i,$ and $D_i$ respectively, where the communication from $S_i$ to $D_i$ is done over $SR_i$--$RD_i$. This point of view reduces our system to $N$ parallel conventional single-relay systems each operating on a single channel, which helps our optimization approach later. Using existing results for the SNR of conventional relay systems \cite{hasna_tw03:alouini}, the overall SNR of the $i$-th subsystem in the AF case is
\begin{equation}
\label{eq:snr_i:af}
\rho_i = \frac{ P_{S_i} P_{R_i} |g_i|^2 |h_i|^2 }{ P_{S_i} |g_i|^2 + P_{R_i} |h_i|^2 + 1 }.
\end{equation}
Similarly, in the ASF case, the corresponding SNR is
\begin{equation}
\label{eq:snr_i:asf}
\rho_{(i)} = \frac{ P_{S_{(i)}} P_{ R_{(i)} } | g_{(i)} |^2 | h_{(i)} |^2 }{ P_{ S_{(i)} } | g_{(i)} |^2 + P_{ R_{(i)} } | h_{(i)} |^2 + 1 }.
\end{equation}

\subsection{Problem formulations}

Depending on the availability of the CSI at the relay and source nodes two different setups are considered: (i) when global CSI is available at both the source node and relay node, i.e., both $S$ and $R$ have knowledge of both $g_i$ and $h_i, \forall i$. (ii) only local CSI is available at the source node, i.e., $S$ knows only $g_i, \forall i$ while $R,$ knows $g_i$ and $h_i, \forall i.$ Thus, in both cases, the transmitter has the knowledge of the channel. It is worth mentioning that for water--filling protocols, it is typical in standards deploying OFDM to provide the CSI to the transmitter \cite{peters_mcom09:heath}.

As mentioned earlier, we consider two different types of relaying: AF and ASF. Thus, a total of four different cases with respect to type of relaying and availability of global CSI can be considered. For these four cases, we study optimal power allocation strategies across $N$ orthogonal links at both $S$ and $R.$ For given finite energy budgets at each node, our goal is to maximize the achievable rate of the network. As can be seen from (\ref{eq:snr_i:af}), the SNR of the link $i,$ i.e., $SR_i$--$RD_i,$ and, thus the capacity of this link depend on both $P_{S_i}$ and $P_{R_i}, \forall i.$ Therefore, we seek the optimal power levels $P_{S_i}$ and $P_{R_i}, \forall i$ that maximize the sum rate of the network. Finding the optimal solution for the ASF relay network can also be motivated in a similar way from (\ref{eq:snr_i:asf}).

The optimal power allocation problems in these four cases can be stated as follows.
\subsubsection{Case I: AF relaying, global CSI at both $S$ and $R$} The goal is to optimally allocate power in both $S$ and $R$ so that the sum capacity of the network is maximized under the individual power constraints at $S$ and $R.$ Thus, we have
\begin{gather}
\label{eq:c1_prb}
\max_{ P_{S_i}, P_{R_i}, \forall i } \,\, \sum_{i=1}^N C_i \\
\label{eq:c1_prb_cons}
\text{ s.t. } \sum_{i=1}^N P_{S_i} \leq P_S, \quad  \sum_{i=1}^N P_{R_i} \leq P_R
\end{gather}
where $C_i$ denotes the capacity of the link $SR_i-RD_i,$ $P_S$ and $P_R$ are the individual power budgets at $S$ and $R,$ respectively.

\subsubsection{Case II: AF relying, global CSI only at $R$} In this case, we study power allocation strategy only at $R.$ This is because, in this case, $S$ knows only $g_i,$ $\forall i$ and, thus, the associated optimal power allocation at $S$ is independent of $h_i, \forall i.$ This means that the optimal power allocation at $S$ is the traditional water--filling rule. Thus, we focus only on optimal power allocation at $R.$

In the first phase, $S$ decides the allocation of its limited power $P_S$ to $N$ orthogonal links using water--filling. For the obtained power allocation, then, $S$ knows the link capacity for $SR_i,$ $\forall i.$ Based on the known link capacity, $S$ decides a data rate $\gamma_i$ for each link $SR_i,$ $\forall i.$

In the second phase, within its limited power $P_R,$ the relay amplifies and forwards the source data received via $N$ links. Particularly, the data transmitted on $SR_i$ should be amplified (and forwarded on $RD_i$) by the relay so that the overall capacity of this link $SR_i$--$RD_i$ is equal to or greater than $\gamma_i.$ Otherwise, channel outage occurs. It is important to notice that the relay, due to its limited power, may not be able to forward all source's data. In such a case, no power should be assigned to the links whose data is not forwarded. In addition, on the links $RD_i$ that the relay decides to forward the source data, it is supposed to put just enough power in order to avoid the channel outage. The goal is, therefore, to forward as much source data as possible. To achieve this goal, we can formulate the following optimization problem
\begin{gather}
\label{eq:c2_prb}
\max\limits_{ P_{R_i}, \forall i } \, \sum_{i=1}^N C_i \\
\label{eq:c2_prb_cons1}
\text{ s.t.} \quad \sum_{i=1}^N P_{R_i} \leq P_R\\
\label{eq:c2_prb_cons2} C_i \in \{ \gamma_i, 0 \}
\end{gather}
where $\gamma_i$ is the data rate decided by $S$ for the link $SR_i,$
and $C_i$ denotes the capacity of the link $SR_i$--$RD_i$
controlled by $P_{R_i}.$

The constraint (\ref{eq:c2_prb_cons2}), as discussed earlier,
is taken into consideration in order to avoid outage on links whose data is forwarded
as well as to avoid power wastage on links whose data is not forwarded.
Also, it is worth stressing that there is no point in allocating extra power on
a link to increase $C_i$ beyond $\gamma_i.$ The choice of $\gamma_i$ will be discussed in Section \ref{sec:opt}, where the optimization problem is solved.

\subsubsection{Case III: ASF relaying, global CSI at both $S$ and $R$} Similar to Case~I, we aim at finding the optimal solution for the power allocation at both $S$ and $R$. However, the difference from Case~I is that the data transmitted on the $i$-th ordered link $SR_{(i)}$ is amplified and forwarded on the $i$-th ordered link $RD_{(i)}.$ Therefore, we can pose the optimization problem as
\begin{gather}
\label{eq:c3_prb}
\max_{ P_{S_{(i)}}, P_{R_{(i)}}, \forall i } \,\, \sum_{i=1}^N C_{(i)} \\
\label{eq:c3_prb_cons}
\text{ s.t. } \sum_{i=1}^N P_{S_{(i)}} \leq P_S, \quad  \sum_{i=1}^N P_{R_{(i)}} \leq P_R
\end{gather}
where $C_{(i)}$ denotes the capacity of the link $SR_{(i)}-RD_{(i)}.$

\subsubsection{Case IV: ASF relying, global CSI only at $R$} In this case, similar to Case II,
the optimization problem can be formulated as
%
\begin{gather}
\label{eq:c4_prb}
\max\limits_{ P_{R_{(i)}}, \forall i } \, \sum_{i=1}^N C_{(i)} \\
\label{eq:c4_prb_cons1}
\text{ s.t.} \quad \sum_{i=1}^N P_{R_{(i)}} \leq P_R, \quad
C_{(i)} \in \{ \gamma_{(i)}, 0 \}
\end{gather}
where $\gamma_{(i)}$ is the data rate decided by $S$ for the link $SR_{(i)},$
and $C_{(i)}$ denotes the capacity of the link $SR_{(i)}$--$RD_{(i)}$
controlled by $P_{R_{(i)}}.$

\section{Optimal rate solution}
\label{sec:opt}

\subsection{Case~I: AF relaying, global CSI at both $S$ and $R$}

In this case, it is assumed that both $S$ and $R$ know $g_i$ and $h_i,$ $\forall i.$ Thus, using (\ref{eq:snr_i:af}), the individual capacity of link $i$ is
$C_i = \log \left( 1 + \rho_i \right).$
Then, the sum capacity of the network is
\begin{equation}
\label{eq:c1_sumcap}
C \triangleq \sum\limits_{i=1}^{N} C_i =
\sum\limits_{i=1}^{N} \log \left( 1 + \frac{ P_{S_i} P_{R_i} |g_i|^2
|h_i|^2 }{ P_{S_i} |g_i|^2 + P_{R_i} |h_i|^2 + 1 } \right).
\end{equation}
Based on (\ref{eq:c1_sumcap}), the optimal power allocation problem (\ref{eq:c1_prb})-\eqref{eq:c1_prb_cons} can be reformulated as
\begin{gather}
\label{eq:c1_opt_prb1}
\max\limits_{ P_{S_i}, P_{R_i}, \forall i } \, \sum_{i=1}^N \log
\left( 1 + \frac{ \alpha_i \beta_i P_{S} P_{R} |g_i|^2 |h_i|^2 }{ \alpha_i P_{S} |g_i|^2 + \beta_i P_{R} |h_i|^2 + 1 } \right) \\
\label{eq:c1_opt_prb1_cons}
\text{ s.t. } \sum\limits_{i=1}^N \alpha_i \leq 1, \quad \sum\limits_{i=1}^N \beta_i \leq 1
\end{gather}
where $\alpha_i$ denotes the ratio of $P_{S_i}$ to $P_S,$ i.e.,
$\alpha_i=P_{S_i} / P_S,$ and similarly $\beta_i = P_{R_i} / P_R.$

Because of constraints on $\alpha$ and $\beta$, solving (\ref{eq:c1_opt_prb1})-\eqref{eq:c1_opt_prb1_cons} directly can be difficult. Therefore, we first find a necessary condition on $\alpha_i$ and $\beta_i$ that significantly helps solving \eqref{eq:c1_opt_prb1}-\eqref{eq:c1_opt_prb1_cons}. For this, we notice that the individual power constraints in (\ref{eq:c1_opt_prb1_cons}) lead to the following necessary condition on the sum of the given individual powers at $S$ and $R$
\begin{equation}
\label{eq:c1_necessary_cond}
\sum_{i=1}^N P_{S_i} + P_{R_i} \leq P_S+ P_R \mbox{~~or~~} \sum_{i=1}^{N} \alpha_i + \sum_{i=1}^{N} \beta_i \tau \le 1 + \tau
\end{equation}
where $\tau$ is the ratio of $P_R$ to $P_S,$ i.e., $\tau = P_R / P_S.$ While this condition is not sufficient,
any conclusion received from enforcing (\ref{eq:c1_necessary_cond}) is
necessary for the optimal solution of (\ref{eq:c1_opt_prb1})-\eqref{eq:c1_opt_prb1_cons}. In
the next theorem, we derive a relationship between
optimal values of $\alpha_i$ and $\beta_i.$

\begin{thry}
\label{thry:1} When $g_i$ and $h_i,$ $\forall i$ are known to both $S$ and  $R,$ the optimal power allocation at $S$ and $R$ satisfies
\begin{equation}
\label{eq:thry1}
\frac{ \beta_i^* }{ \alpha_i^* } \, \tau = \frac{ \alpha_i^* P_S
|g_i|^2 + 1 }{ \beta_i^* P_R |h_i|^2 + 1 }
\end{equation}
\end{thry}
where $\alpha_i^*$ and $\beta_i^*$ denote the optimal values of
$\alpha_i$ and $\beta_i,$ respectively.

\emph{Proof:}
Let $\mathbf{P}_i=[P_{S_i} \; P_{R_i}]$ denote a vector with elements of transmit powers used on subchannel $i.$ Then, it can be shown from (\ref{eq:c1_sumcap}) that for a given $\mathbf{P}_i,$ the sum capacity of the network is a concave function of $\mathbf{P}_i$ since $\partial^2 C / \partial^2 \mathbf{P}_i \leq 0$ for all values of $\mathbf{P}_i \in {\cal R}^2, \forall i.$
Therefore, \eqref{eq:c1_opt_prb1}-\eqref{eq:c1_opt_prb1_cons} is a convex optimization problem with respect to $\mathbf{P}_i,$ $\forall i.$

We now consider the objective (\ref{eq:c1_opt_prb1}) along with the constraint (\ref{eq:c1_necessary_cond}). The objective (\ref{eq:c1_opt_prb1}) can be optimized by using a Lagrangian multiplier method. The associated Lagrange function can be written as
\begin{equation}
\label{eq:c1_objfunc}
{\cal O} = \sum_{i=1}^N \log
\left( 1 + \frac{ \alpha_i \beta_i P_{S} P_{R} |g_i|^2 |h_i|^2 }{ \alpha_i P_{S} |g_i|^2 + \beta_i P_{R} |h_i|^2 + 1 } \right)
+ \lambda \left( 1 + \tau - \sum_{i=1}^N \alpha_i - \sum_{i=1}^N
\beta_i \tau \right)
\end{equation}
where $\lambda$ stands for the Lagrange multiplier.

Taking derivatives of (\ref{eq:c1_objfunc}) with respect to
$\alpha_i$ and $\beta_i,$ $\forall i,$ and equating them to zero,
we obtain that the optimal values of $\alpha_i$ and $\beta_i,$
$\forall i$ can be expressed as:
\begin{gather}
\label{eq:c1_alphai} \alpha_i^* = -\frac{ 2 + \beta_i^* P_R |h_i|^2
}{ 2 P_S |g_i|^2 } + \frac{1}{2}\sqrt{ \beta_i^* \tau \frac{ |h_i|^2
}{ |g_i|^2 }
\left( \beta_i^* \tau \frac{ |h_i|^2 }{ |g_i|^2 } + \frac{4}{ \lambda } \right) }\\
\label{eq:c1_betai} \beta_i^* = -\frac{ 2 + \alpha_i^* P_S |g_i|^2
}{ 2 P_R |h_i|^2 } + \frac{1}{2}\sqrt{ \frac{ \alpha_i^*  |g_i|^2 }{
\tau |h_i|^2 } \left( \alpha_i^* \tau^{-1} \frac{ |g_i|^2 }{ |h_i|^2
} + \frac{4}{ \lambda } \right) }.
\end{gather}

Inserting (\ref{eq:c1_betai}) into (\ref{eq:c1_alphai}), it is revealed
that
\begin{equation}
\label{eq:c1_opt_prb1_albe} \alpha_i^* (\alpha_i^* P_S |g_i|^2 + 1)
= \tau \beta_i^* ( \beta_i^* P_R |h_i|^2 + 1)
\end{equation}
and the claim of Theorem \ref{thry:1} follows straightforwardly.
\hfill $\blacksquare$

Theorem \ref{thry:1} reveals an interesting symmetry for balancing power among individual links. Based on this result, if the optimal $\alpha_i^*$ is known, then the optimal $\beta_i^*$ can be immediately found.

Noticing that (\ref{eq:thry1}) is a necessary condition under which (\ref{eq:c1_opt_prb1}) is optimized under the constraint (\ref{eq:c1_opt_prb1_cons}), we can provide a modified optimization problem only based on $\alpha_i.$
So, the complexity of solving the optimization problem is greatly reduced. Then, using (\ref{eq:c1_sumcap}) and (\ref{eq:thry1}), the sum capacity of the network can be reformulated with respect to only $\alpha_i$ as
\begin{equation}
\label{eq:c1_sumcap2} C = \sum\limits_{i=1}^N \log \left( 1 +
\frac{ \alpha_i P_S |g_i|^2 \left(-1 + A(\alpha_i) \right)/2 }{
\alpha_i P_S |g_i|^2 + \left( 1 + A(\alpha_i) \right)/2 } \right)
\end{equation}
where $A(\alpha_i) = \sqrt{ 1 + 4 \alpha_i P_S |h_i|^2 (\alpha_i P_S
|g_i|^2 + 1) }.$ Then the problem \eqref{eq:c1_opt_prb1}-\eqref{eq:c1_opt_prb1_cons} can be
reformulated as
\begin{gather}
\label{eq:c1_opt_prb2} \max\limits_{ \alpha_i, \forall i } \, \sum_{i=1}^N \log
\left( 1 + \frac{ \alpha_i P_S |g_i|^2 \left(-1 + A(\alpha_i)
\right)/2 }{ \alpha_i P_S |g_i|^2 + \left( 1 + A(\alpha_i) \right)/2
} \right) \\
\label{eq:c1_opt_prb2_cons}
\text{ s.t. } \sum\limits_{i=1}^N \alpha_i \leq 1 \\
\label{eq:c1_opt_prb2_cons2} \sum\limits_{i=1}^N \frac{ -1 +
A(\alpha_i) }{ 2 P_R |h_i|^2 } \leq 1
\end{gather}
where constraint (\ref{eq:c1_opt_prb2_cons2}) is equivalent to
$\sum_{i=1}^N \beta_i \le 1.$ Notice that in this optimization problem, the
objective function and the constraints are all convex functions of
$\alpha_i$. Therefore, traditional efficient numerical convex optimization techniques can
be applied in order to solve it.

\subsection{Case~II: AF relying, global CSI only at $R$}

In this case, as addressed earlier, we assume that in the first phase, $S$ decides a power level $P_{S_i}$ and a data rate $\gamma_i$ for the link $SR_i.$ The goal of the relay in the second phase is to amplify and forward as much source data received over $N$ source--relay links as possible, subject to avoiding outage on subchannels that $R$ decides to forward through. This means that the data received over some links may not be forwarded by the relay, due to its limited available power. The optimization problem in this case is the problem (\ref{eq:c2_prb})--(\ref{eq:c2_prb_cons2}).

For given $\rho_{SR_i}$, we can represent the individual achievable rate of the link $SR_i$--$RD_i$ as
\begin{equation}
\label{eq:c2_capi}
C_i = \log \left( 1 + \frac{ \rho_{SR_i}
\beta_i P_R |h_i|^2 }{ \rho_{SR_i} + \beta_i P_R |h_i|^2 + 1 }
\right).
\end{equation}
The sum rate of the network can therefore be expressed as
\begin{equation}
\label{eq:c2_sumcap}
C \triangleq \sum_{ i =1 } ^{N}
C_i = \sum_{ i =1 } ^{N} \log \left( 1 + \frac{
\rho_{SR_i} \beta_i P_R |h_i|^2 }{ \rho_{SR_i} + \beta_i P_R |h_i|^2
+ 1 } \right)
\end{equation}
where $\beta_i$ is zero for links $RD_i$ that
are not chosen to be amplified.

Based on (\ref{eq:c2_prb}) and (\ref{eq:c2_sumcap}), subject to avoiding outage on likes with $\beta_i \neq 0$, the optimization problem can be reformulated as
\begin{gather}
\label{eq:c2_opt_prb1}
\max\limits_{ \beta_i, \forall i } \, \sum_{ i =1 } ^{N}
 \log \left( 1 + \frac{ \rho_{SR_i} \beta_i P_R |h_i|^2 }{
\rho_{SR_i} + \beta_i P_R |h_i|^2 + 1 } \right) \\
\label{eq:c2_opt_prb1_cons1} \text{ s.t. } \sum_{ i =1
} ^{N} \beta_i \leq 1\\
\label{eq:c2_opt_prb1_cons2} C_i \in \{ {\gamma_i}, 0 \},
\forall  i.
\end{gather}

Due to the discrete constraint (\ref{eq:c2_opt_prb1_cons2}), convex optimization techniques cannot be applied in order to solve \eqref{eq:c2_opt_prb1}--\eqref{eq:c2_opt_prb1_cons2}. When seeking the optimal solution, it may be needed to consider all possible power allocations. The number of different power allocations, however, increases exponentially with $N.$ Therefore, we propose a greedy algorithm which finds near optimal solutions (and in many cases the optimal solution). The idea is to allocate the limited available power in the most efficient way. Particularly, notice that on the relay--destination link $RD_i,$ there is a minimum $\beta_i$ guaranteeing successful source-destination communication. This value of $\beta_i,$ denoted by $\tilde{\beta_i},$ is considered as the cost of communication on this link. This is because $\tilde{\beta_i}$ is proportional to the amount of power spent on this link, if the link is chosen to be amplified and forwarded. Solving
\[
\log \left( 1 + \frac{
\rho_{SR_i} \tilde{\beta_i} P_R |h_i|^2 }{ \rho_{SR_i} + \tilde{\beta_i} P_R |h_i|^2
+ 1 } \right)=\gamma_i,
\]
the value of $\tilde{\beta_i}$ can be found as
\begin{equation}
\label{eq:tild_beta}
\tilde{\beta_i} = \frac{ \left( 2^{\gamma_i}-1 \right) 2^{\gamma_i / \delta_i} }{ \left( 2^{\gamma_i/\delta_i}-2^{\gamma_i} \right) P_R |h_i|^2 }
\end{equation}
where $\delta_i$ denotes the ratio of $\gamma_i$ to the capacity of the link $SR_i.$ While various subchannels can have different $\delta_i$, in this work we assume $\delta_i=\delta, \forall i$ that is the worst--case. In Section \ref{sec:sim}, we observe that even with $\delta_i=\delta, \forall i,$ the achievable sum rate converges at high SNR to the case of Global CSI at both $S$ and $R.$

Consuming the cost $\tilde{\beta_i}$ on link $i,$ in return, the data rate $\gamma_i$ from the source to the destination is obtained. Then, we can define $\eta_i = \gamma_i/\tilde{\beta_i}$ as the efficiency of allocating power to link $i.$ Notice that the larger the value of $\eta_i,$ the better the link $i.$ This is because for the same cost in terms of the power spending, larger data rate $\gamma_i$ is achievable on links that have larger values of $\eta_i.$ So, all links are sorted in the decreasing order of $\eta_i.$ We assign power to the links that have better $\eta_i$ until we run out of power (before $\sum_i \beta_i$ becomes greater than one.)

The greedy algorithm is designed such that the limited power is spent on the best links in terms of the achievable rate. However, this solution may not be optimal because by assigning power according to this greedy algorithm, we may end up with some positive leftover power. In such cases, other strategies that result in zero leftover power, may turn out to achieve a slightly higher rate. Nonetheless, the greedy algorithm is optimal in the sense of using the limited power in the most efficient way and in most cases provides the optimal solution for sum rate maximization problem. In particular, as $N$ grows large (i.e., the effects of the leftover power is negligible), it almost always gives the optimal solution.

Now, we modify this greedy algorithm to avoid any leftover power. The idea is to allow $\delta \in [0,1]$ to be a function of channel statistics. We notice that $\delta$ remains constant as long as channel statistics are constant. To remove the leftover power, we consider the following {$\max-\max$} sum rate optimization problem
\begin{gather}
\label{eq:c2_opt_prb2}
\max\limits_{\delta \in [0,1]} E \left\{ \max\limits_{ \tilde{\beta_i}, \forall i } \, \sum_{ i =1 } ^{N}
 \log \left( 1 + \frac{ \rho_{SR_i} \tilde{\beta_i} P_R |h_i|^2 }{
\rho_{SR_i} + \tilde{\beta_i} P_R |h_i|^2 + 1 } \right)  \right\}
\\
\label{eq:c2_opt_prb2_cons1} \text{ s.t. } \sum_{ i =1
} ^{N} \tilde{\beta_i} \leq 1\\
\label{eq:c2_opt_prb2_cons2} C_i \in \{ {\gamma_i}, 0 \}, \,
\forall  i.
\end{gather}

To solve this optimization problem, for any given value of $\delta \in [0,1],$ we use the aforementioned greedy algorithm to find the maximum sum rate. Then, we find such a $\delta^*$ that provides the maximum sum rate. Please notice that in practice channel statistics do not change fast, so for given channel statistics, the optimal values of $\delta$ can be found off--line and tabulated for later use. As soon as $\delta^*$ is chosen by the network, a single run of our greedy algorithm is needed to find the optimal power allocation per channel realization.

\subsection{Case~III: ASF relaying, global CSI at both $S$ and $R$}
The only difference from Case~I is channel ordering. In other words, data transmitted on $SR_{(i)}$ (the $i$-th ordered link) is amplified and forwarded on $RD_{(i)}.$ The ordering is in the decreasing order of the channel gains $|g_i|^2$ and $|h_i|^2,$ respectively. Therefore, using (\ref{eq:snr_i:asf}), the individual capacity of link $(i)$ can be written as
\begin{equation}
\label{eq:c3_capi}
C_{(i)} = \log \left( 1 + \rho_{(i)} \right).
\end{equation}
where $\rho_{(i)}$ denotes the SNR received via link $SR_{(i)}$--$RD_{(i)}.$
The overall capacity of the network can be defined as
\begin{equation}
\label{eq:c3_sumcap}
C \triangleq \sum_{i=1}^N C_{(i)} = \sum_{i=1}^N \log \left( 1 + \frac{ \alpha_{(i)} \beta_{(i)} P_{S} P_{R} |g_{(i)}|^2
|h_{(i)}|^2 }{ \alpha_{(i)} P_{S} |g_{(i)}|^2 + \beta_{(i)} P_{R} |h_{(i)}|^2 + 1 }  \right)
\end{equation}

Using (\ref{eq:c3_sumcap}), the power allocation optimization problem (\ref{eq:c3_prb})--\eqref{eq:c3_prb_cons} can be represented by the following problem
\begin{gather}
\label{eq:c3_opt_prb1}
\max\limits_{\alpha_{(i)}, \beta_{(i)}, \forall i} \,\, \sum_{i=1}^N \log \left( 1 + \frac{ \alpha_{(i)} \beta_{(i)} P_{S} P_{R} |g_{(i)}|^2 |h_{(i)}|^2 }{ \alpha_{(i)} P_{S} |g_{(i)}|^2 + \beta_{(i)} P_{R} |h_{(i)}|^2 + 1 } \right)\\
\label{eq:c3_opt_prb1_cons}
\text{ s.t. } \sum_{i=1}^N \alpha_{(i)} \leq 1, \quad \sum_{i=1}^N \beta_{(i)} \leq 1.
\end{gather}

Notice that the optimization problem (\ref{eq:c3_opt_prb1})--\eqref{eq:c3_opt_prb1_cons} is mathematically equivalent to the problem (\ref{eq:c1_opt_prb1})--\eqref{eq:c1_opt_prb1_cons} in Case~I. The only difference from \eqref{eq:c1_opt_prb1}--\eqref{eq:c1_opt_prb1_cons} is that the indices of optimal powers $\alpha_{(i)}^*$ and $\beta_{(i)}^*$ in \eqref{eq:c3_opt_prb1}--\eqref{eq:c3_opt_prb1_cons} correspond to the ordered links. Thus, in order to find the optimal solution to \eqref{eq:c3_opt_prb1}--\eqref{eq:c3_opt_prb1_cons}, we follow the same approach as in Case~I by replacing $g_i, h_i, \alpha_i$ and $\beta_i$ with $g_{(i)}, h_{(i)}, \alpha_{(i)}$ and $\beta_{(i)},$ respectively.

\subsection{Case~IV: ASF relying, global CSI only at $R$}
Again, the optimization problem in this case is mathematically identical to the one in Case~II. The only modification needed is to replace $g_i, h_i, \beta_i$ and $\gamma_i$ with  $g_{(i)}, h_{(i)}, \beta_{(i)}$ and $\gamma_{(i)},$ respectively.

\section{Asymptotic analysis with global CSI at both $S$ and $R$ at high SNR}
\label{sec:asym}

In this section, we study the optimization problem of Case~I at high SNR, i.e., when $P_S$ and $P_R$ tend to infinity. Again, we assume $P_R=\tau P_S.$ Thus, for a given $\tau,$ we study the case that $P_S$ tends to infinity. Through this analysis, we discuss the behavior of the optimal solution at high SNR regime. Some interesting observations are discussed.

Using (\ref{eq:snr_i:af}) and (\ref{eq:c1_sumcap2}), when $P_S$ increases, the overall SNR on link $i$ can be given by
\begin{equation}
\label{eq:asym_snr_i:af}
\begin{split}
\lim_{P_S \rightarrow \infty} \rho_i &= \lim_{P_S \rightarrow \infty} \frac{ \alpha_i P_S |g_i|^2 \left(-1 + 2 \alpha_i P_S  |h_i| |g_i| \right)/2 }{
\alpha_i P_S |g_i|^2 + \left( 1 + 2 \alpha_i P_S |h_i| |g_i| \right)/2 } \\
\quad &=  \alpha_i \varphi_i P_S
\end{split}
\end{equation}
where  $\varphi_i \triangleq {|h_i| |g_i|}/ ( 1 + {|h_i| / |g_i|}).$ Then, it can be obtained asymptotically from (\ref{eq:asym_snr_i:af}) that the sum capacity of the network is expressed only in terms of $\alpha_i$ by
\begin{equation}
\label{eq:asym_sumcap:af}
\lim_{P_S \rightarrow \infty} C = \lim_{P_S \rightarrow \infty} \sum_{i=1}^N C_i = \sum_{i=1}^N \log \left( \alpha_i \varphi_i \right) + N \log \left( P_S \right).
\end{equation}

Based on (\ref{eq:c1_opt_prb2}) and (\ref{eq:asym_sumcap:af}), the optimization problem in Case~I can be rewritten at high SNR as
\begin{gather}
\label{eq:c1_asym_optprb}
\max_{\alpha_i, \forall i} \sum_{i=1}^N \log \left( \alpha_i \varphi_i \right) + N \log \left( \frac{ P }{ 1+\tau } \right) \\
\label{eq:c1_asym_optprb_cons1}
\text{ s.t. } \sum_{i=1}^N \alpha_i \leq 1 \\
\label{eq:c1_asym_optprb_cons2}
\frac{1}{\tau} \sum_{i=1}^N \alpha_i  \frac{ |g_i| }{ |h_i| }   \leq 1.
\end{gather}

It is worth noticing that the objective function (\ref{eq:c1_asym_optprb}) is a monotonic function of $\alpha_i,$ $\forall i.$ Using the Lagrange multiplier method, optimal solution to \eqref{eq:c1_asym_optprb}--\eqref{eq:c1_asym_optprb_cons2} can be obtained. Then the Lagrange function can be formulated as
\begin{equation}
\label{eq:asym_objfunc}
{\cal O} = \sum_{i=1}^N \log \left( \alpha_i \varphi_i \right) + N \log \left( \frac{ P}{ 1+\tau } \right) + \lambda_1 \left( 1-\sum_{i=1}^N \alpha_i \right) + \lambda_2 \left( 1-\frac{1}{\tau} \sum_{i=1}^N \alpha_i  \frac{ |g_i| }{ |h_i| }  \right)
\end{equation}
where $\lambda_1$ and $\lambda_2$ are Lagrange multipliers associated with the constraints.
Differentiating (\ref{eq:asym_objfunc}) with respect to $\alpha_i,$ $\forall i$ and equating the results to zero, we obtain the following closed--form expression for the optimal value of $\alpha_i$
\begin{equation}
\label{eq:asym_opt_alp}
\alpha_i^* = \max\left\{0,  \left( \lambda_1 + \frac{\lambda_2}{\tau} \frac{|g_i|}{|h_i|} \right)^{-1}  \right\}.
\end{equation}

In \eqref{eq:asym_opt_alp}, the Lagrange multipliers $\lambda_1$ and $\lambda_2$ are found once for a given set of values $g_i, \forall i$ and $h_i, \forall i$. Therefore, it can be observed that when $\lambda_2>0$, for two subchannels with $h_i=h_j,$ the optimal power allocation solutions put more source power on the weaker subchannel. In other words, if $g_i<g_j$ then $\alpha^*_i > \alpha^*_j,$ and vice versa. This is the inverse of a more traditional water--filling scheme, where only one power constraint exists and the optimal solution puts more power on the stronger subchannel. Similarly, rewriting the above equations in terms of $\beta_i^*,$ we have
\begin{equation}
\label{eq:asym_opt_beta}
\beta_i^* = \max\left\{0,  \left( \lambda'_1 + {\lambda'_2}{\tau} \frac{|h_i|}{|g_i|}  \right)^{-1}  \right\}.
\end{equation}
which means that the behavior of the optimal $\beta^*_i$ is the inverse of water--filling for $\lambda'_2 > 0.$

Furthermore, from Theorem \ref{thry:1}, at high SNR, it is obtained that the optimal solution to the problem \eqref{eq:c1_asym_optprb}--\eqref{eq:c1_asym_optprb_cons2} must satisfy
\begin{equation}
\label{eq:asym_albe}
\lim_{P_S \rightarrow \infty} \alpha_i^* = \tau \beta_i^* \sqrt{ \frac{ |h_i|^2 }{ |g_i|^2 } } \quad \Rightarrow \lim_{P_S \rightarrow \infty} P_S \alpha_i^* = P_R \beta_i^* \sqrt{ \frac{ |h_i|^2 }{ |g_i|^2 } }.
\end{equation}
Notice that $P_S \alpha_i^*$ and $P_R \beta_i^*$ represent the allocated power to links $SR_i$ and $RD_i,$ respectively. It can be concluded then that when $|g_i| > |h_i|$ on link $i,$ the optimal assigned power on link $SR_i$ must be less than the optimal assigned power on link $RD_i.$

\section{Simulations}
\label{sec:sim} Consider a two-hop single relay network where
nodes communicate through $N \in \{ 4, 20 \}$ orthogonal channels.
The source-relay and relay-destination channels are assumed to be
independent Rayliegh flat fading with variance $\sigma^2$. The
total sum power assigned to these $2N$ channels at both the source
and relay is $P = P_S + P_R$. The noise is assumed to be additive
zero-mean white Gaussian with unit variance. Thus, the total
transmit SNR, which is used in our figures, is defined as SNR$= P
\sigma^2.$ Notice that transmit SNRs at $S$ and $R$ can individually be expressed as
$P_S \sigma^2$ and $P_R \sigma^2$, correspondingly.

\subsection{AF relaying}
Assuming that the AF strategy is employed at the relay node,
Fig.~\ref{fig:fig2} shows the achieved sum capacity versus the
total transmit SNR for Case~I when the global CSI is available at
both the source and relay nodes and the proposed optimal power
allocation strategy is used. Here $\tau=1$ (i.e., $P_S=P_R=P/2$).
It can be seen from the figure that the sum capacity increases for
larger $N$. Intuitively, this is because for limited $P_S$ and
$P_R$, larger $N$ provides more chances of obtaining better
source-destination subchannels on which the power allocation can
result in a better sum capacity.

Fig.~\ref{fig:fig3} depicts the results for Case~II, when the
global CSI is available only at the relay and, thus, the power
allocation can only be made at the relay. The sum rate is shown
versus the total transmit SNR for $\tau=1$ and $N = 20$. Three
scenarios for the greedy algorithm are considered. In the first
one, $\delta=0.5$ is fixed for all SNRs, while in the second and
third scenarios, the optimized $\delta$, i.e., $\delta^*$, is
used. The difference between the second and third scenarios is that
water--filling at the source is assumed in one, while equal
power allocation across all subchannels at the source is used in
the other. The proposed greedy algorithm is used in order to
allocate $P_R$ to $N$ orthogonal subchannels. It can be seen from
the figure that, as expected, higher data rates are achieved for
optimal $\delta^*$. Interestingly, the equal power allocation at the source does
not suffer from significant rate loss. Thus, in further examples
for greedy-based power allocation, we consider only equal power
allocation at the source.

In Fig.~\ref{fig:fig4}, the results of the optimal power
allocation in Cases~I and II are compared to each other in terms
of the sum rate plotted versus the total transmit SNR when
$\tau=1$, $N=20$, and optimal $\delta^*$ is used. In agreement
with our analysis, the global optimization in Case~I always
outperforms the other case. Particularly, at sum rate of
$2$~bits/s/Hz, Case~I is shown to obtain $4.5$~dB power gain
compared to Case~II. Here, it is worth mentioning that this
significant power gain is achieved at a modest value of $N$.

The results of the optimal power allocation in Cases~I and II are
also compared to each other in Fig.~\ref{fig:fig5} in terms of the
sum rates when $\tau=0.5$ (i.e., $P_S = 2 P_R$), $N=20$, and
$\delta = \delta^*$. It can be seen from the figure that again, Case~I outperforms Case~II for all values of the total
transmit SNR. However, the performance gap between Cases~I and II decreases as compared to the set up examined in
Fig.~\ref{fig:fig4}. This is because for $\tau \ll 1$ all
source-relay subchannels are likely to have much better conditions
than the relay-destination subchannels, while the sum rate is
mainly affected by the power allocation across the
relay-destination subchannels. In addition, in both figures
(Figs.~\ref{fig:fig4} and Fig.~\ref{fig:fig5}), one can see that
Case~II converges to Case~I at high SNRs. This is because all
subchannels become almost deterministic at high SNR, and
therefore, the global optimization of the power allocation is less
beneficial, while optimization of the power allocation only at the
relay used in Case~II tends to be almost sufficient. This is because Case~II
takes into consideration as many best relay-destination
subchannels as possible within a limited power budget.

Let us now consider $\tau=2$ (i.e., $2 P_S = P_R$). The
corresponding sum rates is shown versus the total transmit SNR for
both Cases~I and II in Fig. \ref{fig:fig6} for $N=20$ and $\delta
= \delta^*$. It can be seen from this figure that Case~I is better
than Case~II for all values of SNR. Interestingly, comparing all cases tested, i.e., $\tau=0.5$
(Fig.~\ref{fig:fig5}), $\tau=1$ (Fig.~\ref{fig:fig4}), and
$\tau=2$ (Fig. \ref{fig:fig6}), the global optimal power
allocation is most beneficial when $\tau=1$.

\subsection{ASF relaying}
Fig.~\ref{fig:fig7} depicts the sum rate versus the total transmit
SNR for Case~III when the global CSI is available at both the
source and relay nodes and the optimal power allocation is performed for ASF relaying. For this figure, $N \in \{4, 20\}$ and $\tau=1$. Case~I is also depicted for comparison purposes. It
can be observed from this figure that Case~III outperforms Case~I, as expected.

Optimizing $\delta$ and the power allocation only at the relay in Case~IV, the sum rate is shown in
Fig.~\ref{fig:fig8} versus the total transmit SNR for $N=20$ and
$\tau=1$. The sum rate curves for Cases~I, II, and III are also
depicted for comparison. It is confirmed in
Fig.~\ref{fig:fig8} that Case~III always outperforms all other
cases including Case~IV for the same reasons that have been
explained while comparing Cases~I and II. Interestingly, it can
also be observed in this figure that Case~IV is superior to the
global optimization of Case~I for moderate and high SNRs.

\section{Conclusions}
\label{sec:con}

Two--hop AF and ASF relay networks, consisting of single source, destination, and relay nodes with limited individual
power constraints at the source and relay are considered. Optimal power
allocation strategies across $N$ orthogonal subchannels between nodes are studied,
depending on the availability of CSI at both the source and relay or only at the relay.

When source has global CSI knowledge, through a symmetry property proved for the optimal power allocation, the optimization problem is solved. Via an asymptotic analysis, it is found that the global power optimization assigns more power on the weaker subchannels, which is the inverse of the traditional water--filing. When only the relay has global CSI knowledge, a greedy algorithm maximizing the achievable sum rate is proposed. For this, the actual data rate at the source is optimized and the minimum powers on subchannels are found at the relay in order to guarantee successful source-destination communication. It is also shown numerically that the optimal power allocation performed only at the relay can outperform the global power optimization scheme at moderate and large SNRs if simple subchannel sorting capabilities are added at the relay.

\linespread{1.6}

\begin{figure}[hbt]
\centering
\includegraphics[width=12.5cm]{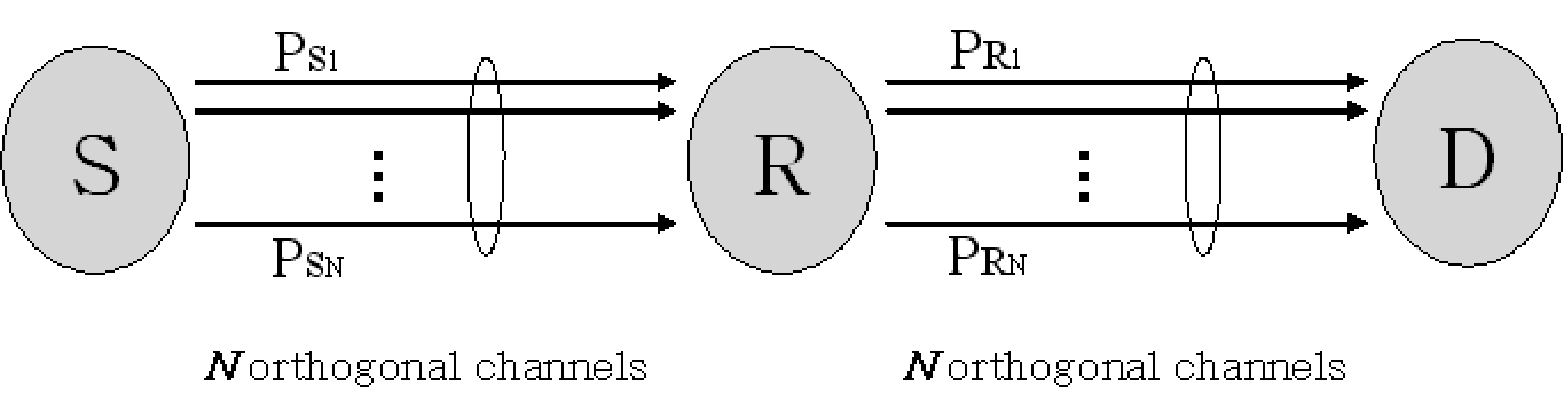}
\caption{Block diagram of a two-hop single relay wireless network
with $N$ source-relay and $N$ relay-destination orthogonal channels.}
\label{fig:fig1}
\end{figure}

\begin{figure}[hbt]
\centering
\includegraphics[width=11.5cm]{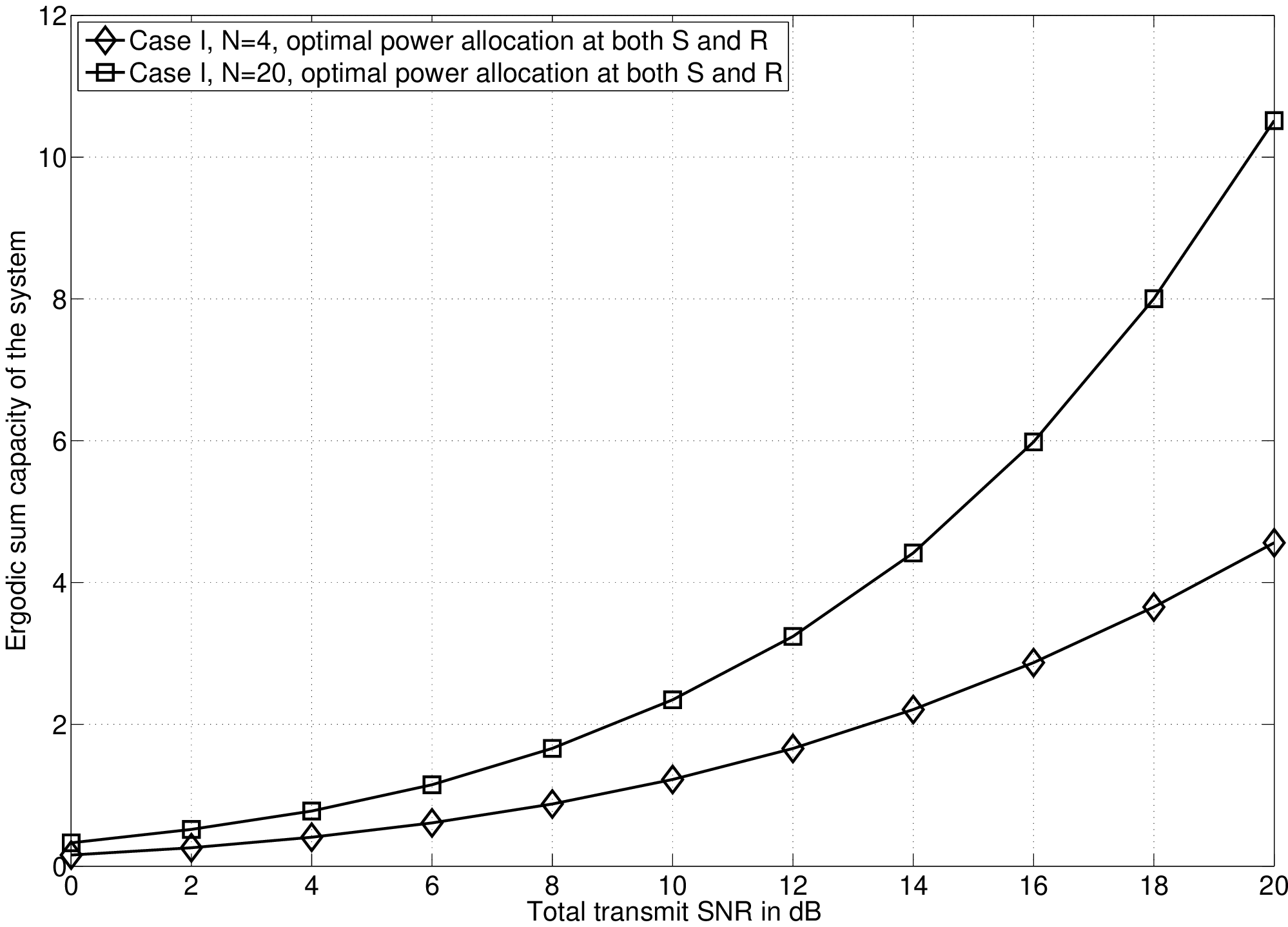}
\caption{The sum capacity versus the total transmit SNR for Case~I: the optimal power allocation at both the source and the relay for $N\in\{4, 20\}$ and $\tau=1.$} \label{fig:fig2}
\end{figure}

\begin{figure}[hbt]
\centering
\includegraphics[width=11.5cm]{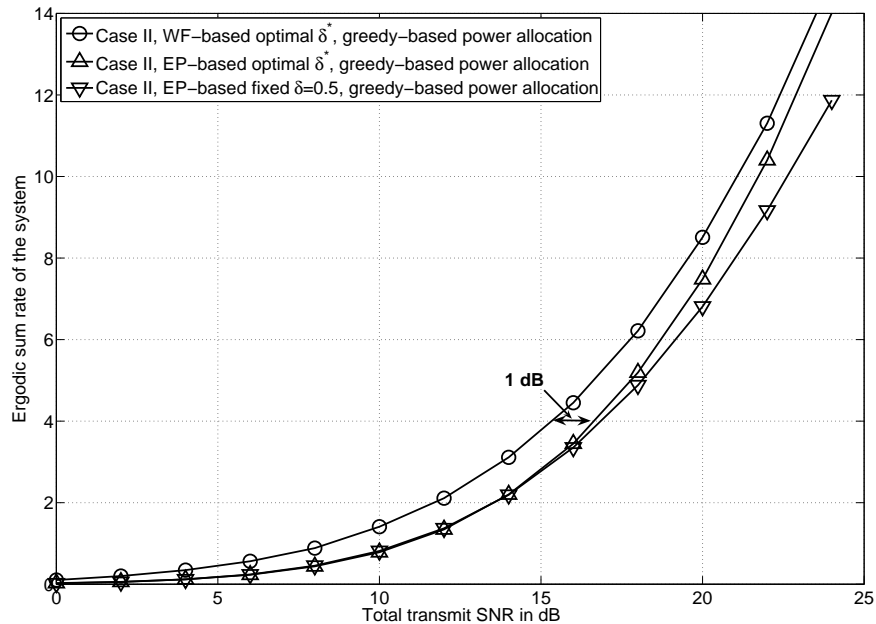}
\caption{The sum rate versus the total transmit SNR using the proposed greedy algorithm for $N=20$ and $\tau=1.$ Three cases of $\delta$ are examined: (i) optimal $\delta^*$ with the water--filling at the source (WF-based optimal $\delta^*$); (ii) optimal $\delta^*$ with the equal power allocation across $N$ orthogonal channels at the source (EP-based optimal $\delta^*$); (iii) fixed $\delta=0.5$ with the equal power allocation across $N$ orthogonal channels at the source (EP-based fixed $\delta=0.5$).} \label{fig:fig3}
\end{figure}

\begin{figure}[hbt]
\centering
\includegraphics[width=11.5cm]{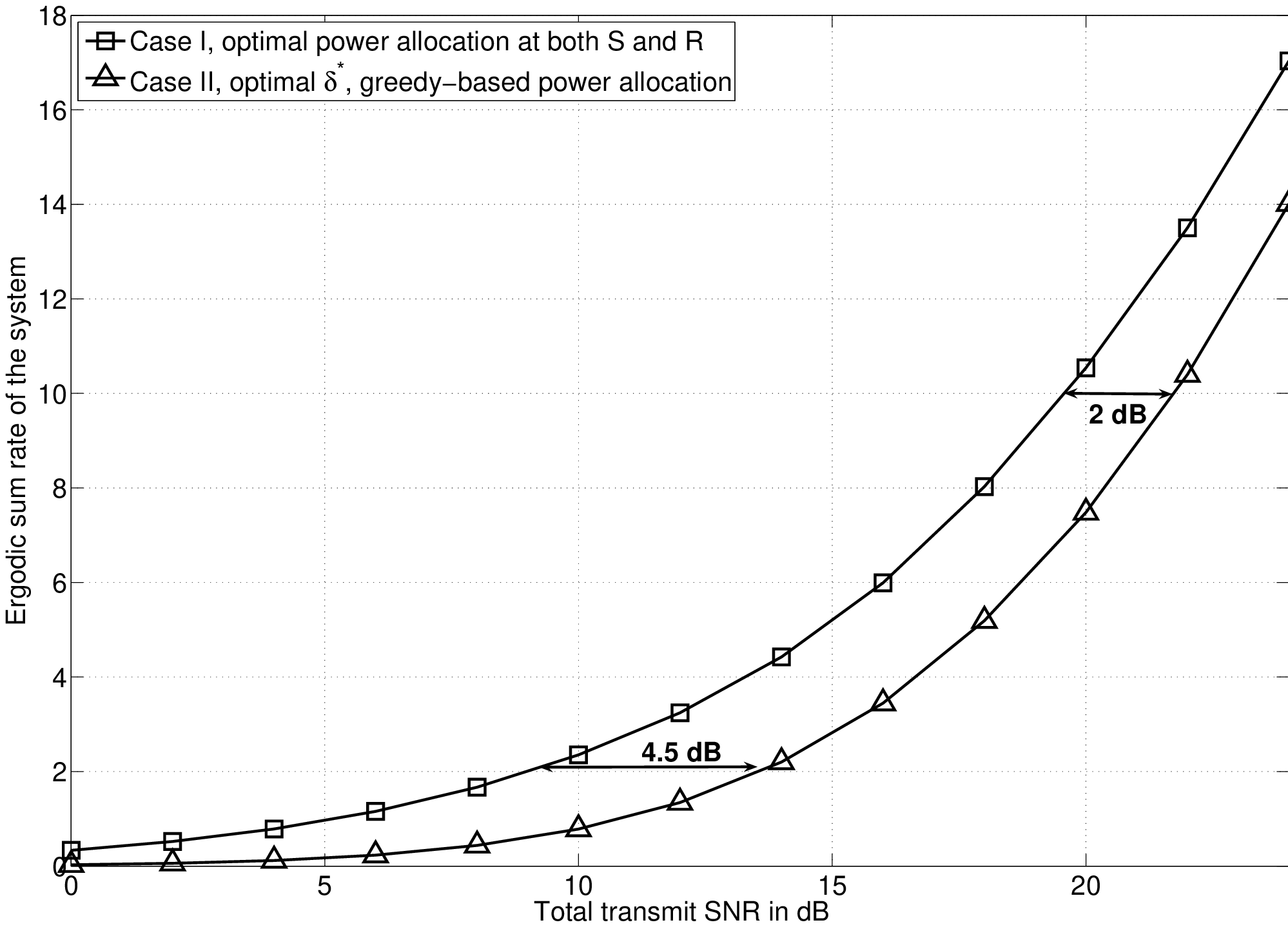}
\caption{Comparison of the sum rates for Cases~I and II: $N=20,$ $\tau=1,$ and $\delta = \delta^*$ for equal power allocation across $N$ orthogonal channels at the source.} \label{fig:fig4}
\end{figure}

\begin{figure}[hbt]
\centering
\includegraphics[width=11.5cm]{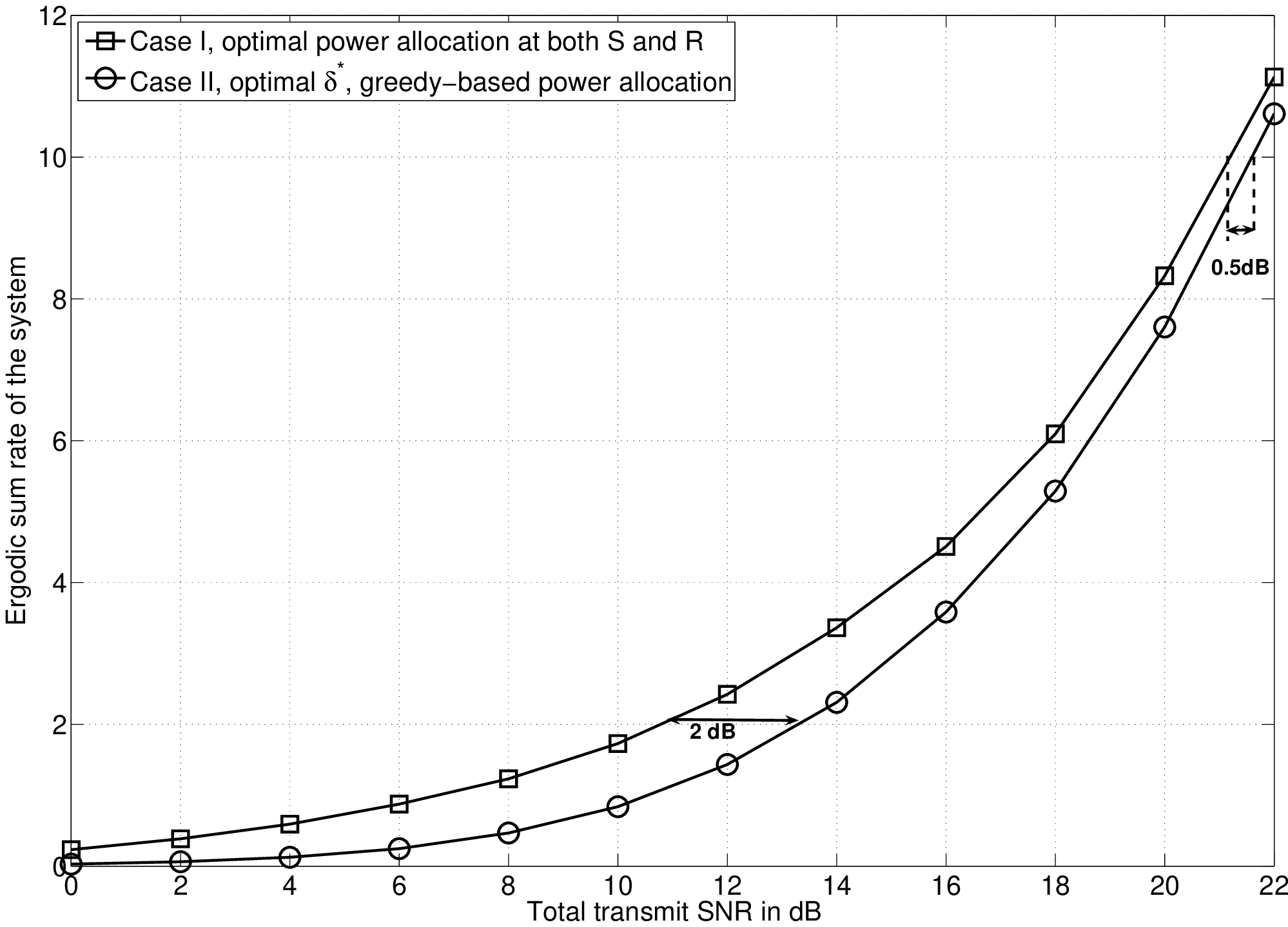}
\caption{Comparison of the sum rates for Cases~I and II: $N=20, \tau=0.5,$ and $\delta = \delta^*$ for equal power allocation across $N$ orthogonal channels at the source.} \label{fig:fig5}
\end{figure}

\begin{figure}[hbt]
\centering
\includegraphics[width=11.5cm]{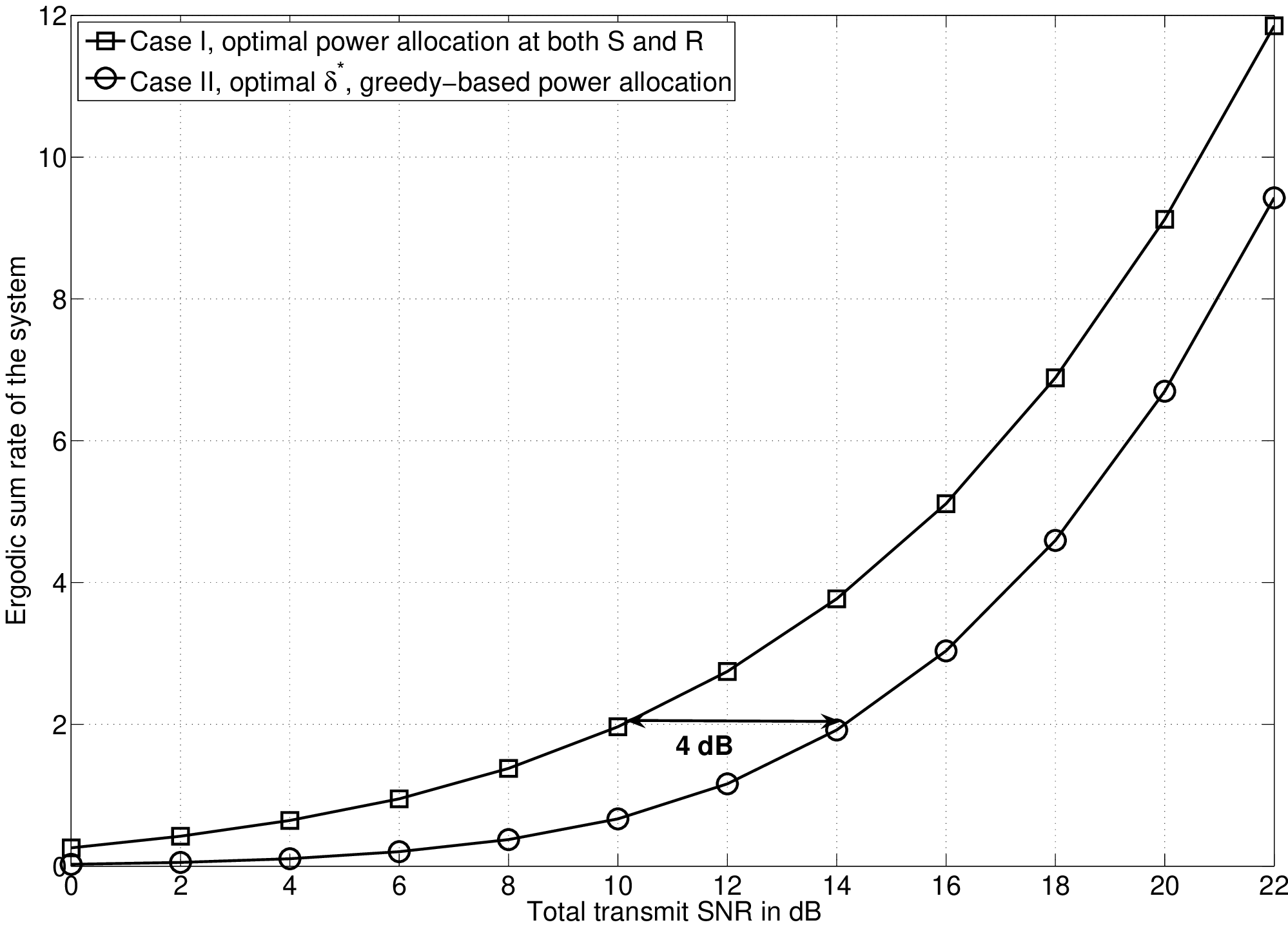}
\caption{Comparison of the sum rates for Cases~I and II: $N=20, \tau=2,$ and $\delta = \delta^*$ for equal power allocation across $N$ orthogonal channels at the source.} \label{fig:fig6}
\end{figure}

\begin{figure}[hbt]
\centering
\includegraphics[width=11.5cm]{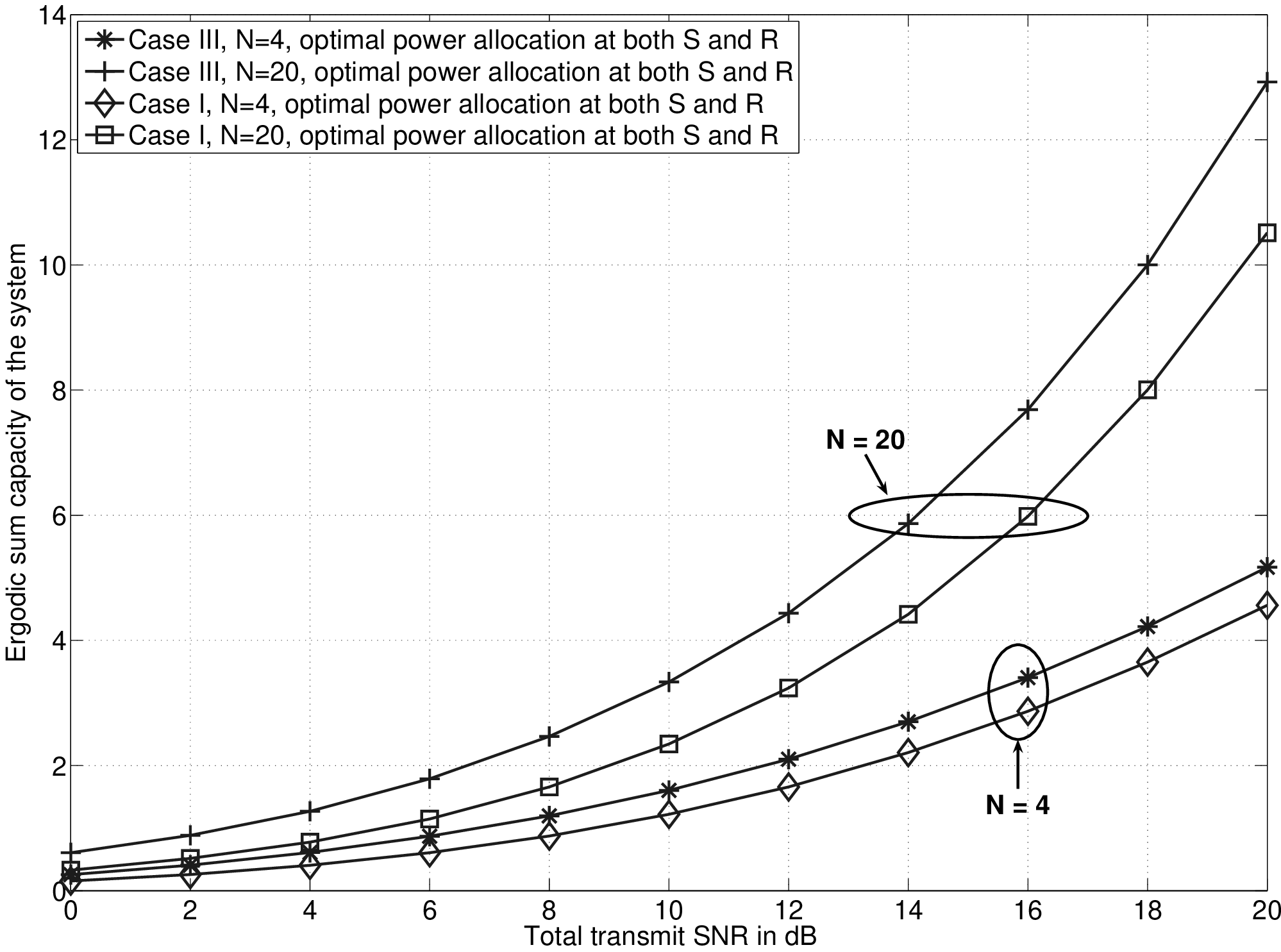}
\caption{The sum capacity versus the total transmit SNR: Case~III of the optimal power allocation at both the source and the relay for $N\in\{4, 20\},$ and $\tau=1.$ Case~I is depicted for comparison.}
\label{fig:fig7}
\end{figure}

\begin{figure}[hbt]
\centering
\includegraphics[width=11.5cm]{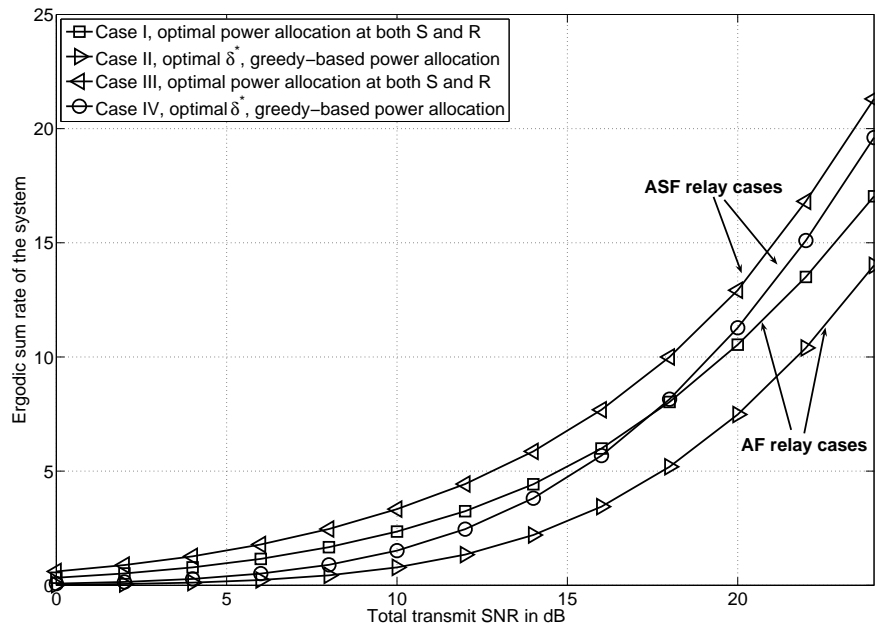}
\caption{The sum rate versus the total transmit SNR using the proposed greedy algorithm in Case~IV for $N=20, \tau=1,$ $\delta = \delta^*$ for equal power allocation across $N$ orthogonal channels at the source. Cases~I, II, and III are depicted for comparison.}
\label{fig:fig8}
\end{figure}

\end{document}